\documentclass[onecolumn,showpacs]{revtex4}

\usepackage{graphicx}
\usepackage{dcolumn}
\usepackage{bm}

\input epsf

\begin{document}

\title{\Large Generalized Cosmic Chaplygin Gas Model with or without Interaction}

\author{\bf Writambhara
Chakraborty$^1$\footnote{writam1@yahoo.co.in}, Ujjal
Debnath$^2$\footnote{ujjaldebnath@yahoo.com} and Subenoy
Chakraborty$^3$\footnote{subenoyc@yahoo.co.in} }

\affiliation{$^1$Department of Mathematics, New Alipore College,
New Alipore, Kolkata- 700 053, India.\\
$^2$Department of Mathematics, Bengal Engineering and Science
University, Shibpur, Howrah-711 103, India.\\
$^3$ Department of Mathematics, Jadavpur University, Kolkata-700
032, India.}

\date{\today}

\begin{abstract}
Recently developed Generalized Cosmic Chaplygin gas (GCCG) is
studied as an unified model of dark matter and dark energy. To
explain the recent accelerating phase, the Universe is assumed to
have a mixture of radiation and GCCG. The mixture is considered
for without or with interaction. Solutions are obtained for
various choices of the parameters and trajectories in the plane
of the statefinder parameters and presented graphically. For
particular choice of interaction parameter, we have shown the
role of statefinder parameters in various cases for the evolution
of the Universe.
\end{abstract}


\maketitle

\section{\normalsize\bf{Introduction}}

Recent observations of type Ia Supernovae indicate that the
expansion of the Universe is accelerating [1-5] and lead to the
search for a new type of matter which violates the strong energy
condition, i.e., $\rho+3p<0$. The matter responsible for this
condition to be satisfied at some stage of evolution of the
universe is referred to as {\it dark energy} [6 - 8]. Several
candidates to present dark energy have been suggested with
observations: the cosmological constant [7, 9], quintessence [10,
11], phantom [12, 13], braneworld models [14], pure Chaplygin gas
model [15], generalized Chaplygin gas (GCG) model [16, 17],
modified Chaplygin gas (MCG) model [19, 20]. In the GCG and MCG
approach dark energy and dark matter can be unified by using an
exotic equation of state (EOS). Interesting feature of MCG (or
GCG) EOS is that it shows radiation era (or dust era) in the
past while a $\Lambda$CDM model in the future.\\

In 2003, P. F. Gonz$\acute{a}$lez-Diaz [21] have introduced the
generalized cosmic Chaplygin gas (GCCG) model in such a way that
the resulting models can be made stable and free from unphysical
behaviours even when the vacuum fluid satisfies the phantom
energy condition. The EOS of this model is
\begin{equation}
p=-{\rho}^{-\alpha}\left[ C + (\rho^{1+\alpha}-C)^{-w}\right]
\end{equation}
where $C=\frac{A}{1+w}-1$ with $A$ a constant which can take on
both positive and negative values and $-l<w<0$, $l$ being a
positive definite constant which can take on values larger than
unity. The EOS reduces to that of current Chaplygin unified
models for dark matter and dark energy in the limit $w\rightarrow
0$ and satisfies the conditions: (i) it becomes a de Sitter fluid
at late time and when $w=-1$, (ii) it reduces to $p=w \rho$ in the
limit that the Chaplygin parameter $A\rightarrow 0$, (iii) it
also reduces to the EOS of current Chaplygin unified dark matter
models at high energy density and (iv) the evolution of density
perturbations derived from the chosen EOS becomes free from the
pathological behaviour of the matter power spectrum for
physically reasonable values of the involved parameters at late
time. This EOS shows dust era in the past and $\Lambda$CDM in the
future.\\

Since models trying to provide a description of the cosmic
acceleration are proliferating, there exists the problem of
discriminating between the various contenders. To this aim Sahni
et al [22] proposed a pair of parameters $\{r,s\}$, called {\it
statefinder} parameters. In fact trajectories in the $\{r,s\}$
plane corresponding to different cosmological models demonstrate
qualitatively different behaviour. The above statefinder
diagnostic pair has the following form:

\begin{equation}
r=\frac{\dddot{a}}{aH^{3}}~~~~\text{and}~~~~s=\frac{r-1}{3\left(q-\frac{1}{2}\right)}
\end{equation}

where $H\left(=\frac{\dot{a}}{a}\right)$ and
$q~\left(=-\frac{a\ddot{a}}{\dot{a}^{2}}\right)$ are the Hubble
parameter and the deceleration parameter respectively. The new
feature of the statefinder is that it involves the third
derivative of the cosmological radius. These parameters are
dimensionless and allow us to characterize the properties of dark
energy. Trajectories in the $\{r,s\}$ plane corresponding to
different cosmological models, for example $\Lambda$CDM model
diagrams correspond to the fixed point $s=0,~r=1$.\\

In this paper, we consider the Universe is filled with the
mixture of radiation and GCCG in section II. We perform a
statefinder diagnostic to this model without and with interaction
in different cases in sections III and IV respectively. From
statefinder parameters we have shown graphically that the
universe starts from radiation era instead of dust era. Different
phases of the evolution of the universe have been shown
graphically. With interaction case, the model goes from radiation
to $\Lambda$CDM era only and without interaction case the model
goes from radiation to $\Lambda$CDM and further from $\Lambda$CDM
to phantom era and then back to $\Lambda$CDM. The paper ends with a short discussion in section V.\\

\section{\normalsize\bf{Mixture of GCCG and radiation}}

The metric of a spatially flat isotropic and homogeneous Universe
in FRW model is,

\begin{equation}
ds^{2}=dt^{2}-a^{2}(t)\left[dr^{2}+r^{2}(d\theta^{2}+sin^{2}\theta
d\phi^{2})\right]
\end{equation}

where $a(t)$ is the scale factor.\\

The Einstein field equations are (choosing $8\pi G=c=1$)

\begin{equation}
3\frac{\dot{a}^{2}}{a^{2}}=\rho_{tot}
\end{equation}
and
\begin{equation}
6\frac{\ddot{a}}{a}=-(\rho_{tot}+3p_{tot})
\end{equation}

The energy conservation equation ($T_{\mu;\nu}^{\nu}=0$) is
\begin{equation}
\dot{\rho}_{tot}+3\frac{\dot{a}}{a}(\rho_{tot}+p_{tot})=0
\end{equation}

where, $\rho_{tot}$ and $p_{tot}$ are the total energy density and
the pressure of the Universe, given by,
\begin{equation}
\rho_{tot}=\rho+\rho_{r}
\end{equation}
and
\begin{equation}
p_{tot}=p+p_{r}
\end{equation}
with $\rho$ and $p$ are respectively the energy density and
pressure due to the GCCG satisfying the EOS (1) and $\rho_{r}$ and
$p_{r}$ are the energy density and the pressure corresponding to
the radiation fluid with EOS,
\begin{equation}
p_{r}=\gamma \rho_{r}
\end{equation}
where $\gamma=\frac{1}{3}$.\\

Since GCCG can explain the evolution of the Universe starting from
dust era to $\Lambda$CDM, considering the mixture of GCCG with
radiation would make it possible to explain the evolution of the
Universe from radiation to $\Lambda$CDM.\\

\section{\normalsize\bf{Without Interaction}}

In this case GCCG and the radiation fluid are conserved
separately. Conservation equation (6) yields,
\begin{equation}
\dot{\rho}+3\frac{\dot{a}}{a}(\rho+p)=0
\end{equation}
and
\begin{equation}
\dot{\rho_{r}}+3\frac{\dot{a}}{a}(\rho_{r}+p_{r})=0
\end{equation}

From equations (1), (9), (10), (11) we have
\begin{equation}
\rho=\left[C +
\left(1+\frac{B}{a^{3(1+\alpha)(1+w)}}\right)^{\frac{1}{1+w}}\right]^{\frac{1}{1+\alpha}}
\end{equation}
and
\begin{equation}
\rho_{r}=\rho_{0}~ a^{-3(1+\gamma)}
\end{equation}

For the two component fluids, equation (2) takes the following
forms:
\begin{equation}
r=1+\frac{9}{2(\rho+\rho_{r})}\left[\frac{\partial
p}{\partial\rho}(\rho+p)+\frac{\partial
p_{r}}{\partial\rho_{r}}(\rho_{r}+p_{r})\right]
\end{equation}
and
\begin{equation}
s=\frac{1}{(p+p_{r})}\left[\frac{\partial
p}{\partial\rho}(\rho+p)+\frac{\partial
p_{r}}{\partial\rho_{r}}(\rho_{r}+p_{r})\right]
\end{equation}

Also the deceleration parameter $q$ has the form:
\begin{equation}
q=-\frac{\ddot{a}}{aH^{2}}=\frac{1}{2}+\frac{3}{2}\left(\frac{p+p_{r}}{\rho+\rho_{r}}\right)
\end{equation}

Now substituting $u=\rho^{1+\alpha}, ~ y=\frac{\rho_{r}}{\rho}$,
equation (14) and (15) can be written as,
\begin{equation}
r=1+\frac{9}{2(1+y)}\left[ \left(1 -
\frac{C}{u}-\frac{(u-C)^{-w}}{u} \right) \left\{ \frac{\alpha
C}{u}+\frac{\alpha}{u}(u-C)^{-w}+w(1+\alpha)(u-C)^{-w-1}\right\}+\gamma
(1+\gamma) y \right]
\end{equation}
and
\begin{equation}
s=\frac{2(r-1)(1+y)}{9\left[ \gamma y -
\frac{C}{u}-\frac{(u-C)^{-w}}{u}\right]}
\end{equation}

Normalizing the parameters we have shown the graphical
representation of the $\{r, s\}$ parameters in figure 1. From the
figure we have seen that the universe starts from radiation era
$(r=3,~s>0)$ via dust stage $(2.3<r<2.4,~s\rightarrow\pm\infty)$
to $\Lambda$CDM ($r=1,~s=0$) model for choices of $ C=1, B=1,
\alpha=1, w=-2, \rho_{0}=1$.

\begin{figure}
\includegraphics[height=3in]{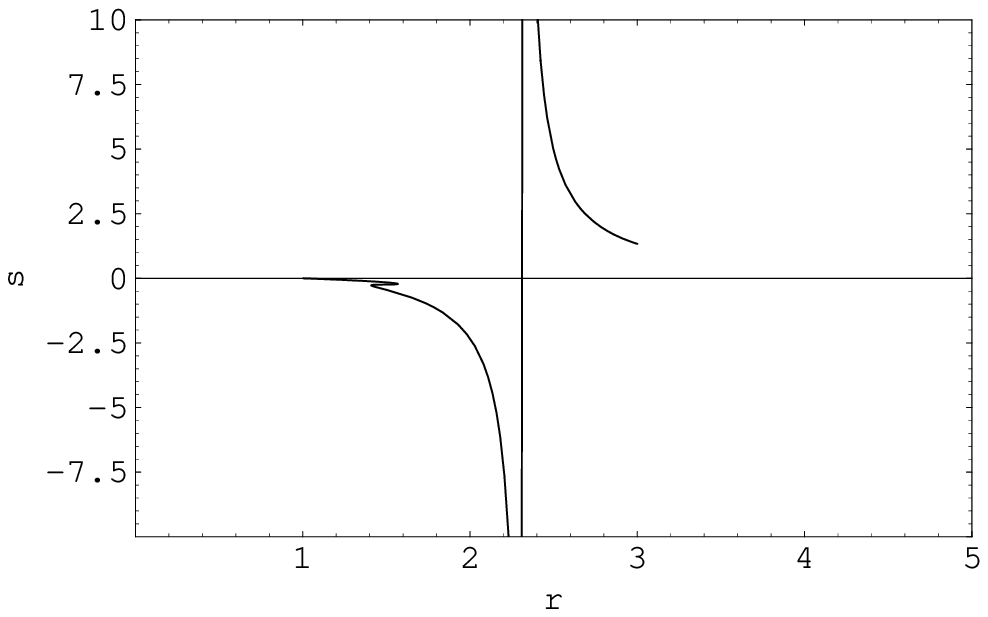}

Fig.1 \vspace{3mm}

\vspace{5mm} Fig. 1 shows the variation of $s$
 against $r$ (eqs. (17) and (18)) for $ C=1, B=1, \alpha=1, w=-2, \rho_{0}=1$. \hspace{14cm} \vspace{4mm}

\end{figure}

\section{\normalsize\bf{With Interaction}}

We consider the GCCG interacting with radiation fluid through an
energy exchange between them. The equations of motion can be
written as,
\begin{equation}
\dot{\rho}+3\frac{\dot{a}}{a}(\rho+p)=-3H\delta
\end{equation}
and
\begin{equation}
\dot{\rho_{r}}+3\frac{\dot{a}}{a}(\rho_{r}+p_{r})=3H\delta
\end{equation}
where $\delta$ is a coupling function.\\

Let us choose,
\begin{equation} \delta=\epsilon~
\frac{(\rho^{1+\alpha}-C)^{-w}}{\rho^{\alpha}}
\end{equation}

Now equation (19) together with equation (1) gives,
\begin{equation}
\rho=\left[C + \left(1-\epsilon+ B
a^{3-(1+\alpha)(1+w)}\right)^{\frac{1}{1+w}}\right]^{\frac{1}{1+\alpha}}
\end{equation}
Also equations (9), (20) and (22) give
\begin{equation}
\rho_{r}=\rho_{0}~ a^{-3(1+\gamma)}+ 3~ \epsilon ~a
^{-3(1+\gamma)}I
\end{equation}
with
\begin{equation}
I=-\frac{1}{3 B (1+\alpha)}\int
\frac{dx}{(C+x)^{\frac{\alpha}{1+\alpha}}}\left\{\frac{x^{1+w}+\epsilon-1}{B}\right\}^{-\frac{1+\gamma}{(1+w)(1+\alpha)}-1}
\end{equation}
and
\begin{equation}
x=\left[ 1-\epsilon + B a^{-3(1+w)(1+\alpha)}
\right]^{\frac{1}{1+w}}
\end{equation}
From (22), we see that if $\epsilon=0$, i.e., $\delta=0$, then
the expression (22) reduces to the expression (12).\\\\

Now for the two component interacting fluids with equations of
motion (19) and (20), the $\{r, s\}$ parameters read:
\begin{equation}
r=1+\frac{9}{2(\rho+\rho_{r})}\left[\frac{\partial
p}{\partial\rho}(\rho+p+\delta)+\frac{\partial
p_{r}}{\partial\rho_{r}}(\rho_{r}+p_{r}-\delta)\right]
\end{equation}
and
\begin{equation}
s=\frac{2(r-1)(\rho+\rho_{r})}{9(p+p_{r})}
\end{equation}

Also the deceleration parameter $q$ has the form:
\begin{equation}
q=\frac{1}{2}\left(1+3\frac{p+p_{r}}{\rho+\rho_{r}}\right)
\end{equation}

Now substituting $u=\rho^{1+\alpha}, ~ y=\frac{\rho_{r}}{\rho}$,
equation (14) and (15) can be written as,
\begin{equation}
r=1+\frac{9}{2(1+y)}\left[\frac{\partial p}{\partial
\rho}\left(1+\frac{p}{\rho}+\frac{\delta}{\rho}\right)+\gamma\left\{(1+\gamma)y-\frac{\delta}{\rho}\right\}
\right]
\end{equation}
and
\begin{equation}
s=\frac{2(r-1)(1+y)}{9\left( \frac{p}{\rho}+\gamma y \right)}
\end{equation}
where,
$$u=\left[C + \left(1-\epsilon+ B
a^{3-(1+\alpha)(1+w)}\right)^{\frac{1}{1+w}}\right] $$
$$ y=\frac{\rho_{0}}{\rho}~ a^{-3(1+\gamma)}+ 3~\frac{ \epsilon}{\rho} ~a
^{-3(1+\gamma)}I$$
$$\frac{p}{\rho}=-\frac{1}{u}\left\{C+(u-C)^{-w}\right\} $$
$$\frac{\delta}{\rho}=\epsilon \frac{(u-C)^{-w}}{u} $$
and $$ \frac{\partial p}{\partial \rho}=\frac{\alpha
C}{u}+\frac{\alpha}{u}(u-C)^{-w}+w(1+\alpha)(u-C)^{-w-1}$$\\

Now we find the exact solution for the $\{r, s\}$ parameters for
the following particular choices of $w$:\\

(i) If $-\frac{(1+\gamma)}{(1+w)(1+\alpha)}-1=0$, i.e.,
$w=\frac{-2-\gamma-\alpha}{1+\alpha}$, equation (23) can be
written as
\begin{equation}
\rho_{r}=\rho_{0}~a^{-3(1+\gamma)}-\frac{\epsilon}{B}a^{-3(1+\gamma)}
\rho
\end{equation}
as $I=-\frac{1}{3B}(c+x)^{\frac{1}{1+\alpha}}$. Normalizing the
parameters, the corresponding statefinder parameters are given in
figure 2. From the figure we have seen that the universe starts
from radiation era $(r=3,~s>0)$ via dust stage
$(2.5<r<2.6,~s\rightarrow\pm\infty)$ to $\Lambda$CDM ($r=1,~s=0$)
model and further from $\Lambda$CDM to phantom era ($r<1,~s>0$)
and then back to $\Lambda$CDM for choices of $ C= 1$, $B=1,
\alpha=1, w=-\frac{5}{3}, \rho_{0}=1, \epsilon=\frac{1}{2}$.\\

\begin{figure}
\includegraphics[height=3in]{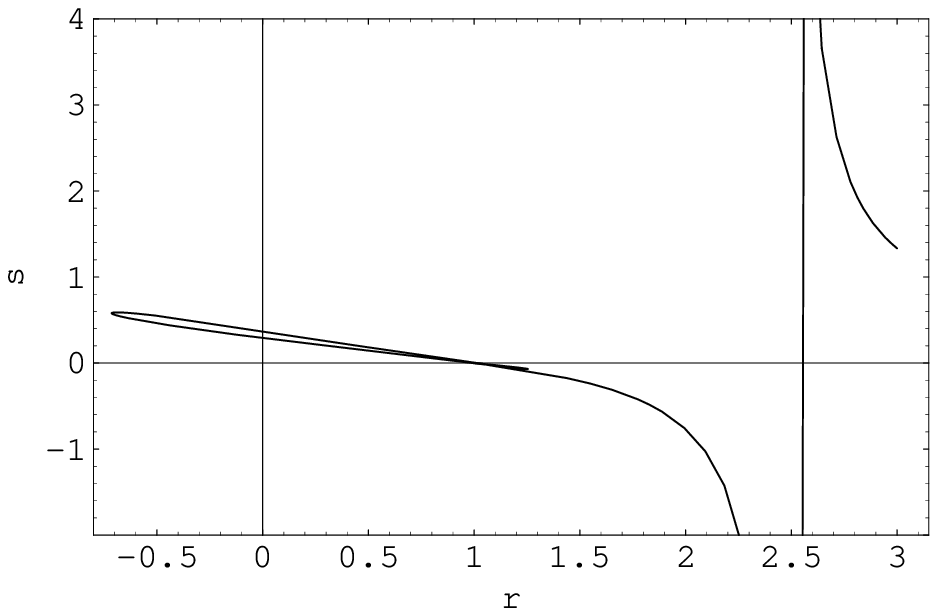}

Fig.2 \vspace{3mm}

\vspace{5mm} Fig. 2 shows the variation of $s$  against $r$ (case
(i) ) for $ C= 1$, $B=1, \alpha=1, w=-\frac{5}{3}, \rho_{0}=1,
\epsilon=\frac{1}{2}$. \hspace{14cm} \vspace{4mm}

\end{figure}

\begin{figure}
\includegraphics[height=3in]{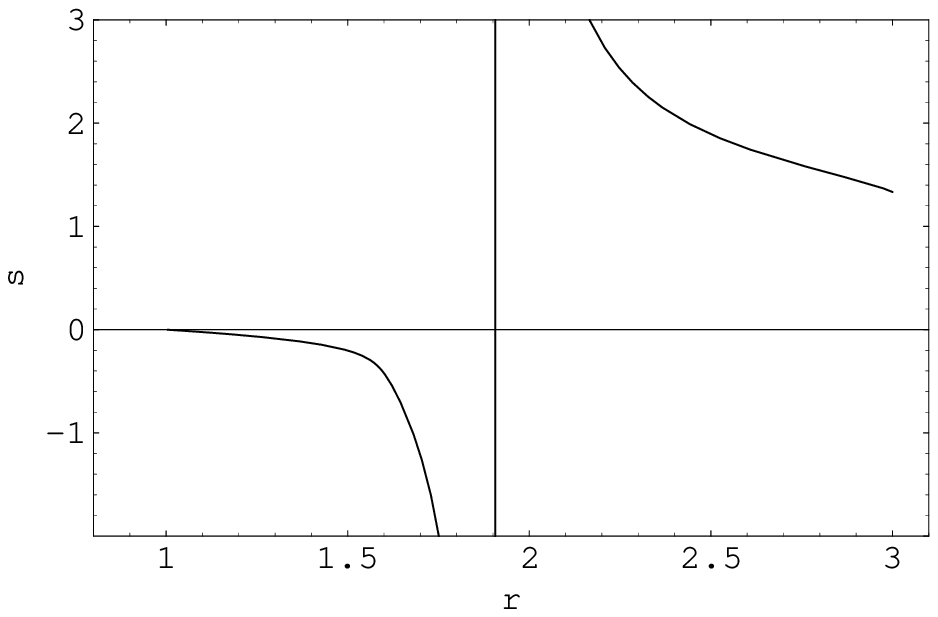}

Fig.3 \vspace{3mm}

\vspace{5mm} Fig. 3 shows the variation of $s$  against $r$ (case
(ii) ) for $ C=1$, $B=1, \alpha=1, w=-\frac{4}{3}, \rho_{0}=1,
\epsilon=\frac{1}{2}$. \hspace{14cm} \vspace{4mm}

\end{figure}

(ii) If $-\frac{(1+\gamma)}{(1+w)(1+\alpha)}-1=1$, i.e.,
$w=\frac{-3-\gamma-2\alpha}{2(1+\alpha)}$, equation (23) can be
written as
\begin{equation}
\rho_{r}=\rho_{0}~a^{-3(1+\gamma)}-\frac{\epsilon(\epsilon-1)}{B^{2}}
a^{-3(1+\gamma)}-\frac{\epsilon
a^{-3(1+\gamma)}}{B^{2}(1+\alpha)(2+w)
C^{\frac{\alpha}{1+\alpha}}} x^{2+w} ~_{2}F_{1}[2+w,
\frac{\alpha}{1+\alpha}, 3+w, -\frac{x}{C}]
\end{equation}
Normalizing the parameters, the corresponding statefinder
parameters are given in figure 3. From the figure we have seen
that the universe starts from radiation era $(r=3,~s>0)$ via dust
stage $(1.9<r<2,~s\rightarrow\pm\infty)$ to $\Lambda$CDM
($r=1,~s=0$) model for choices of $ C=1$,
$B=1, \alpha=1, w=-\frac{4}{3}, \rho_{0}=1, \epsilon=\frac{1}{2}$.\\

(iii) If $w=-2$, equation (23) can be written as
\begin{equation}
\rho_{r}=\rho_{0}~a^{-3(1+\gamma)}-\frac{
\epsilon}{(1+2\alpha-\gamma)} \frac{
a^{-3(1+\gamma)}}{x^{\frac{1+2\alpha-\gamma}{(1+\alpha)}}}
\frac{B^{-\frac{1+\gamma}{(1+\alpha)}}}{
C^{\frac{\alpha}{1+\alpha}}}
 Appell F_{1} \left[
\frac{1+2\alpha-\gamma}{(1+\alpha)}, \frac{\alpha}{1+\alpha},
\frac{\alpha-\gamma}{(1+\alpha)},
\frac{2+3\alpha-\gamma}{(1+\alpha)}, -\frac{x}{C}, x-x \epsilon
\right]
\end{equation}
Normalizing the parameters, the corresponding statefinder
parameters are given in figure 4. From the figure we have seen
that the universe starts from radiation era $(r=3,~s>0)$ via dust
stage $(2.3<r<2.5,~s\rightarrow\pm\infty)$ to $\Lambda$CDM
($r=1,~s=0$) model and further from $\Lambda$CDM to phantom era
($r<1,~s>0$) and then back to $\Lambda$CDM for choices of $ C=1$,
$B=1, \alpha=1, \rho_{0}=1, \epsilon=\frac{1}{2}$.

\begin{figure}
\includegraphics[height=3in]{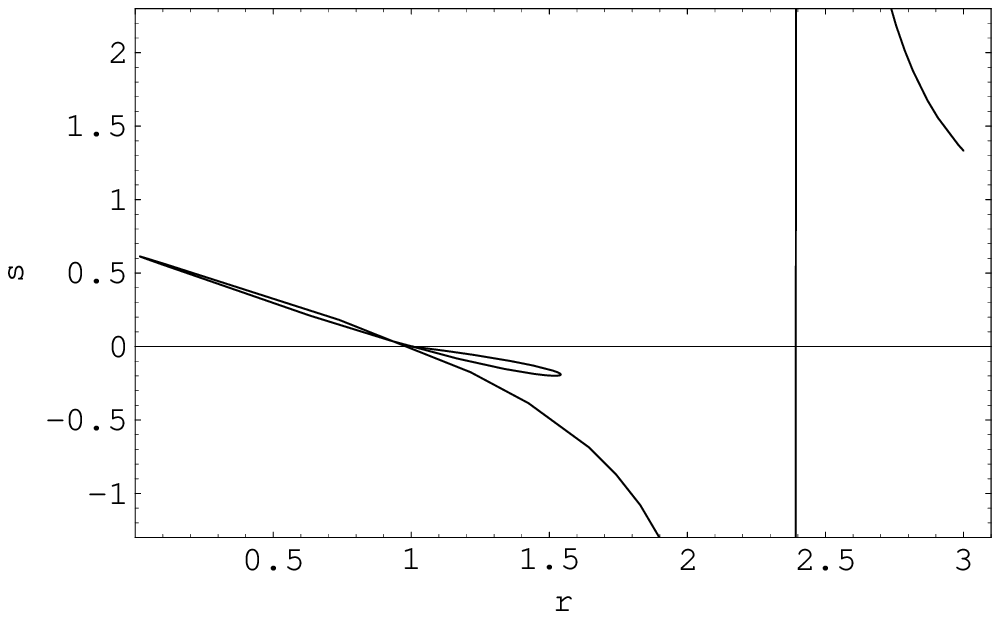}

Fig.4 \vspace{3mm}

\vspace{5mm} Fig. 4 shows the variation of $s$  against $r$ (case
(iii) ) for $ C=1$, $B=1, \alpha=1, \rho_{0}=1,
\epsilon=\frac{1}{2}$. \hspace{14cm} \vspace{4mm}

\end{figure}

\section{\normalsize\bf{Discussion}}

In this work, we have considered the matter in our Universe as a
mixture of the GCCG and radiation as GCCG can explain the
evolution of the Universe from dust era to $\Lambda$CDM. These
gases are taken both as non-interacting and interacting mixture.
In the first case we have considered a non-interacting model and
plotted the $\{r, s\}$ parameters. As expected this model
represents the evolution of the Universe from radiation era to
$\Lambda$CDM with a discontinuity at dust. In the second case the
interaction term is chosen in a very typical form to solve the
corresponding conservation equations analytically. Also the
statefinder parameters are evaluated for various choices of
parameters and the trajectories in the $\{r,s\}$ plane are
plotted to characterize different phases of the Universe. These
trajectories show discontinuity at same $r$ in the neighbourhood
of $r=2$ and have peculiar behaviour around $r=1$. The $\{r,s\}$
curves have two branches on two sides of the asymptote. The
branch on the right hand side of the asymptote corresponds to
decelerating phase before (or up to) dust era, while the left
hand side branch has a transition from decelerating phase upto
$\Lambda$CDM era. Some peculiarity has been shown in figures 2
and 4 around $r=1$. In these two cases, the model goes further
from $\Lambda$CDM to phantom era and then back to $\Lambda$CDM.
Moreover, in figure 4, there is further transition from
$\Lambda$CDM to decelerating phase and then then again back to
$\Lambda$CDM. Thus we can conclude that the present model
describes a number of transitions from decelerating to
accelerating phase and vice-versa.\\\\

{\bf Acknowledgement:}\\

The authors are thankful to IUCAA, India for warm hospitality
where part of the work was carried out. Also UD is thankful to
UGC, Govt. of India for providing research project grant (No. 32-157/2006(SR)).\\

{\bf References:}\\
\\
$[1]$ S. J. Perlmutter et al, {\it Bull. Am. Astron. Soc.} {\bf
29} 1351 (1997).\\
$[2]$ S. J. Perlmutter et al, {\it Astrophys. J.} {\bf 517} 565
(1999).\\
$[3]$ A. G. Riess et al, {\it Astron. J.} {\bf 116} 1009 (1998).\\
$[4]$ P. Garnavich et al, {\it Astrophys. J.} {\bf 493} L53
(1998).\\
$[5]$ B. P. Schmidt et al, {\it Astrophys. J.} {\bf 507} 46 (1998).\\
$[6]$ V. Sahni and A. A. Starobinsky, {\it Int. J. Mod. Phys. A}
{\bf 9} 373 (2000).\\
$[7]$ P. J. E. Peebles and B. Ratra, {\it Rev. Mod. Phys.} {\bf
75} 559 (2003).\\
$[8]$ T. Padmanabhan, {\it Phys. Rept.} {\bf 380} 235 (2003).\\
$[9]$ S. M. Carrol, {\it Living Rev. Rel.} {\bf 4} 1 (2001).\\
$[10$] R. R. Caldwell, R. Dave and P. J. Steinhardt, {\it Phys.
Rev. Lett.} {\bf 80} 1582 (1998).\\
$[11]$ V. Sahni and L. Wang, {\it Phys. Rev. D} {\bf 62} 103507
(2000).\\
$[12]$ R. R. Caldwell, {\it Phys. Lett. B} {\bf 545} 2 (2002).\\
$[13]$ S. M. Carrol, M. Hoffman and M. Trodden, {\it Phys. Rev.
D} {\bf 68} 023509 (2003).\\
$[14]$ U. Alam, V. Sahni, {\it astro-ph}/ 0209443.\\
$[15]$ A. Kamenshchik, U. Moschella and V. Pasquier, {\it Phys.
Lett. B} {\bf 511} 265 (2001).\\
$[16]$ V. Gorini, A. Kamenshchik and U. Moschella, {\it Phys.
Rev. D} {\bf 67} 063509 (2003).\\
$[17]$ U. Alam, V. Sahni , T. D. Saini and
A.A. Starobinsky, {\it Mon. Not. Roy. Astron. Soc.} {\bf 344}, 1057 (2003).\\
$[18]$ M. C. Bento, O. Bertolami and A. A. Sen, {\it Phys. Rev. D}
{66} 043507 (2002).\\
$[19]$ H. B. Benaoum, {\it hep-th}/0205140.\\
$[20]$ U. Debnath, A. Banerjee and S. Chakraborty, {\it Class.
Quantum Grav.} {\bf 21} 5609 (2004).\\
$[21]$ P. F. Gonz$\acute{a}$lez-Diaz, {\it Phys. Rev. D} {\bf 68}
021303 (R), (2003).\\
$[22]$ V. Sahni, T. D. Saini, A. A. Starobinsky and U. Alam,
{\it JETP Lett.} {\bf 77} 201 (2003).\\

\end{document}